\documentclass[12pt,a4paper,notitlepage]{article}
\usepackage{graphicx}
\usepackage{amsmath}
\usepackage{amsfonts}
\usepackage{amssymb}
\begin{document}





\title{Pressure Dependence of the Barrier Height in Tunnel $n$-$GaAs/Au$ Junctions}
\author{
E.M. Dizhur$^{1}$, A.Ya. Shul'man$^{2}$, I.N. Kotel'nikov$^{2}$, A.N.
Voronovsky$^{1}$\\
$^{1}$Institute for High Pressure Physics of the RAS,\\ Troitsk 142090, Moscow Reg., Russia\\
$^{2}$Institute of Radioengineering and Electronics of the RAS,\\ Moscow 103907, Russia}
\date{}
\maketitle

\begin{abstract}
The theory of tunnel current-voltage ($I$-$V$) characteristics of
metal-semiconductor junctions based on the self-consistent solution of Poisson
equation allows to get the Schottky-barrier height and the charged impurity
concentration directly from the tunneling data. This approach was applied to
the analysis of the low temperature experiments on tunneling under pressure up
to $3\,$GPa in a piston-cylinder gauge. Here we present the barrier height
versus pressure for heavily doped $n$-$GaAs$($Te$)$/Au$ ($N_{e}\sim5-
7\cdot10^{18}$ cm$^{-3}$) tunnel junctions and compare the obtained pressure
dependence of the Schottky barrier with known behavior of the band gap under
pressure taking into account the influence of the $L$- and $X$-valleys and DX centers.
\end{abstract}

The knowledge of doping level and the potential barrier height at the
interface as well as their dependence on pressure is important for studies
of the semiconductor structures where the surface band bending region
is essential. Our previous works\ \cite{ICPS90,JETP92DX}\ showed that it
is possible to carry out qualitative low-temperature tunneling spectroscopy
experiments at pressure up to $3$ Gpa using stand-alone high-pressure cell.
The aim of this work is to extend this technique to quantitative study of
band bending region in heavily-doped semiconductors under high pressure.

The pinning of the Fermi level that determines the magnitude of Schottky
barrier may be attributed to metal-induced gap states \cite{SSI95}. In the
particular case of the $Au$\ contact to $n$-$GaAs$ ($100$) plane the barrier
formation was studied by photoemission spectroscopy\ on $N_{e}=5\cdot10^{18}$
cm$^{-3}$ doped material \cite{PRB92}. The barrier height $\Phi_{s}=1$ eV was
obtained at $Au$\ coverage beginning from $1$ monolayer. This value coincides
with the results obtained in \cite{FTP87} by means of tunneling spectroscopy
for the same or less doped case, however, the junctions with higher doping
level indicated decreasing the barrier height. Nevertheless, too little
is known about the barrier in the case of heavily doped $GaAs$, especially
under pressure exceeding approximately $1.5\,$GPa, when the electron states
related to above lying \ $L$- and/or $X$-minima\ might overlap with the states
in $\Gamma$-minimum of the conduction band occupied by the electrons.

Obtaining the suitable data experimentally is rather a difficult task in the
case of heavily-doped semiconductors, because the free carrier tunneling across
the barrier prevents from implementation of the usual techniques like
capacitance-voltage ($C$-$V$) or barrier photo-e.m.f. measurements
\cite{FTP87}. The use of tunneling current itself instead seems to be a
promising solution.

The tunneling measurements under pressure started as early as in 1963
\cite{PR63}, and since then were sporadically used for investigations of
$p$-$n$ tunnel diodes, Schottky junctions, quantum wells and so on. The only
known example of such an approach to the tunnel Schottky junction has been
demonstrated in the work \cite{PRB72} where $Pb/n$-$GaAs$ ($N_{e}\simeq
5\cdot10^{18}$cm$^{-3}$) structures under pressures up to $1.7$ GPa were
studied. However, the barrier height as a function of pressure was not
determined.

One should note that the interpretation of the results obtained from the
tunneling measurements crucially depends on the validity of the model
describing the dependence of the tunneling current on the bias voltage.

In this paper we present the results of experimental study of tunneling in
Schottky junction $Au/n$-$GaAs$($Te$) at a doping level exceeding $N_{e}%
\sim5\cdot10^{18}$cm$^{-3}$ under pressure up to $3$ GPa.

The pressure up to $3$ GPa was generated at room temperature in a stand-alone
high-pressure cell of a piston-cylinder type \cite{JETP79} filled with $40\%$
transformer oil and $60\%$ pentane mixture as a pressure transmitting medium.
After slow cooling down to low temperatures the $I(V)$ curves were measured
within the accuracy $7\,1/2$ digit DC. The actual pressure was evaluated by
the change of the critical temperature $T_{c}$ of $Sn$ wire \cite{PR58} placed
{\it in situ} using the expression: $\Delta T_{c}=-0.495P+0.039P^{2}$ (the
pressure $P$ in GPa) within $0.02$ GPa accuracy.

The tunnel junctions with intimate $Au$-$GaAs$ interface were made by a method
described in \cite{FTT85} and the stability of the samples under pressure was
ensured by hf sputtering of a $SiO$ film approximately $200$ nm thick on the
$GaAs$ surface coinciding with the ($100$) plane. This film was deposited
around a gold electrode $0.25$ and $1$ mm in diameter. Next, two copper
contacts to the gold electrode were deposited by evaporation (see the inset in
Fig. 2). Using this contact formation technology, the $I$-$V$ characteristics
were reproducible and returned to the initial curves after releasing the
pressure. Parameters of the junctions under investigation are presented in
Table 1 along with two junctions studied in \cite{PRB72} and included for
comparison.

In Fig. 1 the results of measurements are presented in terms of junction
conductance $\sigma(V)=dI/dV$ at $4.2$ K versus bias voltage $V$ for
different pressures $P$. At $P\lesssim2$ GPa the known drop of the tunnel
conductance with pressure takes place usually ascribed mainly to an increase
of the barrier height. However, at $P\geq2$ GPa drastic changes appear in the
shape and magnitude of $\sigma(V)$ curves. This effect is clearly
manifested by the curve at $P=2.5$ GPa for the sample $1$ on the left panel
and the curve $2.2b$ for the sample $2$.

It is worth mentioning the observation of some instabilities and other
unobvious behavior of $\sigma(V)$ at $P\simeq2$ GPa. For example, some
kind of switching can occur. The curve $2.2b$ on the middle panel of Fig. 1 was
obtained in a couple of days later than $2.2a$ and differs from the latter not
only by value but also by the overall shape. The pressure cell was warmed up
to ambient temperature but not unloaded between the runs. This is neither the
pressure leakage since we control the pressure by measuring superconductivity
of $Sn$ wire placed {\it in situ} , nor the sample damage since after releasing
pressure the characteristics of the samples returned to the original ones.
Previously the similar behavior was noted for the tunnel $p$-$n$ $GaAs$
junctions at $1.8$ GPa \cite{JETP84}, but there the switching took seconds. In
the case of the sample $1$ at $2.15$ GPa a small hysteresis in $\sigma(V)$
curves appeared as the direction of the bias sweeping was changed and,
besides, $\sigma(V)$ decreased by about $10\%$ during heating from $4.2$
K to $77$ K instead of usual increase.

The determination of the bulk electron density $N_{e}$ and the surface barrier
height $\Phi_{s}$ from $I$-$V$ characteristics of Schottky-barrier tunnel
junctions  was suggested and experimentally tested on $n-GaAs/Au$ contacts
in \cite{FTT85,FTP87}. This approach was based on the expression for tunnel
current including quasi-classical formula for the barrier transparency and
the exact first integral of the Poisson equation for Coulomb potential in
the semiconductor. The present investigation required further modification
of the theoretical description to account for possible decrease of the free
electron density in $\Gamma$-valley of the conduction band under hydrostatic
pressure at high doping level of $n$-$GaAs$ substrates.

This effect is known to exist in the case of $Si-$doped $GaAs$\ at $0\leq
P\leq1.5$ GPa and is referred to as an appearance of DX-centers \cite{DDD90M,
JAP90R}. The properties of $GaAs$($Te$) are much worse understood in
this respect. In \cite{PSSa90K,SST91Te} the measurements of Shubnikov-
de Haas effect in the bulk samples with $N_{e}=7\cdot10^{18}$cm$^{-3}$
revealed  no change up to $P=1.5$ GPa. Therefore, the absence of emersion
of DX($Te$) states resonant with $\Gamma$-conduction band which could capture
free electrons, was claimed. On the other hand, high-pressure
photoluminescence study \cite{SST91PL} revealed the appearance of a hole
recombination center in the $1.5$-$2$ GPa range. Our previous low-temperature
high-pressure experiments with tunnel $n$-$GaAs$($Te$)$/Au$ junctions combined
with X-ray microprobe analysis have also shown such effects (usually ascribed
to the presence of active DX centers) as twice lower electron density in the
bulk than that of the charged impurity density in the depletion layer of
Schottky barrier and weak traces of persistent photoconductivity in tunnel
$I$-$V$ characteristics \cite{ICPS90,JETP92DX}. It is known for very
long time that the free electron density in $Te$-doped $GaAs$ becomes less
than the impurity atom density beginning from $N_{e}\gtrsim5\cdot10^{18}$
cm$^{-3}$ and never exceeds $10^{19}$cm$^{-3}$ being up to two times less than
total $Te$ concentration \cite{Nasled73}. Nevertheless, no firm information is
available up to now about the energy position of DX($Te$) level in $GaAs$,
its dependence on pressure, free carrier density, etc. \cite{SST91Te,
DDD90B}.

Beside above-mentioned problems, at high pressure limit of the present
investigations a possibility existed to encounter the free electron transition
from $\Gamma$-valley to subsidiary $L$- and/or $X$-valleys of the conduction
band.

To take into account the possible difference between bulk free electron
density in $\Gamma$-valley and charged impurity density in the depletion layer
within the scope of the theoretical consideration, on one hand, and to avoid
unjustified detailed elaboration of DX($Te$)-center model, on the other hand,
the following assumptions were accepted:

1. Only $\Gamma$-valley free electrons take part in tunneling process forming
the tunnel current since these electrons have the smallest effective mass and
the lowest barrier height.

2. In the high electric field of the Schottky barrier all electrons possibly
captured by any kind of traps should be released due to the tunneling
ionization. As a result, the density of the positive charge in
the depletion layer may be inhomogenious and may differ from the bulk free
electron density. According to the investigations of deep impurity-center
tunneling ionization by DC and terahertz range electric field \cite{Gan97}
the characteristic value of the field should be of order of $10^{5}$ V/cm.
Due to the exponential field dependence of the ionization rate deep levels
are supposed to be totally emptied in the region with electric field of
comparative or higher magnitude. The electric fields in Schottky barrier can
even exceed such values. Thus, the simplifying suggestion has been accepted
that in the depletion layer the border between the region with partially
ionized centers and totally ionized ones is very sharp and may be described
as a step-like discontinuity in the charged impurity distribution.


The data treatment procedure is based on the expression for
the density of tunnel current that may be written as:
\begin{equation}
I\left(  V,T\right)  =\frac{em_{c}}{\pi\hslash^{3}}\int\limits_{0}^{\infty
}dE\left[  f\left(  E,T\right)  -f\left(  E+eV,T\right)  \right]
\int\limits_{0}^{\varepsilon}dE_{\Vert}D\left(  E,E_{\Vert},V\right)
.\label{Eq1}%
\end{equation}
Here $f(E,T)$ is the Fermi distribution function with temperature $T$, $E$ is
the electron energy, $E_{\Vert}=(\hslash\mathbf{k}_{\Vert})^{2}/2m_{c}$,
$\mathbf{k}_{\Vert}$ is the electron wave vector along the junction plane,
$m_{c}$ is the electron effective mass at the bottom of $\Gamma$-valley of the
conduction band,$\ V$ is the bias voltage. The quasi-classical expression for
the barrier transparency $D$ in Franz two-band approximation can be presented
in the form
\begin{equation}
D\left(  \varepsilon,\varepsilon_{\Vert},V\right)  =\exp\left(  -\frac
{2\sqrt{2m_{c}\mu_{F}^{0}}L_{s}}{\hslash}\int_{\varepsilon}^{\varphi_{b}}%
\frac{d\psi}{d\psi/dx}\sqrt{(\psi-\varepsilon)[1-(\psi-\varepsilon
)/\varepsilon_{\Gamma}]+\varepsilon_{\Vert}}\right)  ,.\label{Eq2}%
\end{equation}
where all variables denoted by small symbols in the integrand are in
dimensionless form, i.e. the total energy $\varepsilon$\ and the barrier
potential $\psi$ are normalized by the characteristic energy $\mu_{F}%
^{0}=\hslash^{2}k_{F}{}^{2}/2m_{c}$,
and spatial coordinate $x$ is normalized by the characteristic length $L_{s}%
=\sqrt{\kappa\mu_{F}^{0}/8\pi e^{2}N_{e}}$. The other quantities denote:
$k_{F}=(3\pi^{2}N_{e})^{1/3}$, $\mu$\ is the Fermi energy of electron plasma
in the semiconductor, $\varphi_{b}=\varphi_{s}+\mu-eV/$ $\mu_{F}^{0}$ is the
band-bending height at the semiconductor-metal interface, $\varepsilon
_{\Gamma}$ is the band gap for $\Gamma$-valley, and $\kappa$ is the
low-frequency dielectric constant of the semiconductor.

It is necessary to note the importance of exact integration over
$\varepsilon_{\Vert}$ in Eq. (\ref{Eq1}) because the Schottky barrier becomes
very thin at such high electron densities. It was also necessary to include
the exchange potential in highly degenerate electron gas into the consideration
for correct  description of the shape of the barrier potential and,
therefore, the tunnel $I$-$V$ characteristics over the bias region of the
order of Fermi energy $\mu\approx150\div200$ meV in our samples. The details
of exact calculation of Eq. (\ref{Eq2}) for the transparency with account for
the self-consistent solution of the Kohn-Sham and Poisson equations may be
found in \cite{SZ76}. Here the further generalization was made to account for
the two-band energy spectrum of electrons.

Due to the complicated situation with the origin of the distinction between
the free electron and impurity atom concentrations discussed above, the
following simplified approach was accepted. In the bulk of the semiconductor
the neutrality condition requires that the electron density $N_{e}%
$\ should be equal to the positive charge density $N_{+}.$ This constant value
is denoted by $N_{0}$ to distinguish it from the variable densities $N_{e}%
(x)$.and $N_{+}(x)$ depending on the spatial coordinate $x$ (see Fig. 2).
Inside the depletion layer the high  barrier electric field may give rise to
additional ionization of impurity atoms by the tunneling process and hence to
the increase of the positive charge density.

Thus, beside the $N_{0}$ and the surface potential $\Phi_{s}$
the maximum density of ionized impurity atoms in the depletion layer
and the electric field $E_{cr}$ below which the density of charged ions is
equal to that of free electrons have been chosen as fitting parameters.
In other words, the Fermi energy $\mu$ of electrons in $\Gamma$-valley
and the density ratio $N_{+}/N_{0}$\ of ionized impurity atoms in the
high-field region of the barrier were also considered as free parameters to
be determined by fitting procedure. This is the main difference of
the present approach from that developed in \cite{FTP87}.

The experimental data were fitted to the model using known pressure
dependence of the energy gap $E_{\Gamma}=1.514+10.8\cdot10^{-2}P;$
($P$ is in GPa, $E_{\Gamma}$ is in eV) \cite{PRB42}, the dielectric constant
$d\kappa/dP=-0.0881/GPa$ \cite{PRB41} and assuming the pressure dependent
electron effective mass at the bottom of $\Gamma$-valley in the form $m_{c}%
(P)=m_{c}(0)(1+\Delta E_{\Gamma}(P)/E_{\Gamma})$. A good coincidence of the
measured and calculated $I(V)$ curves was attained using a minimum least
square four-parameter fitting . Even the differential resistance turned out
well fitted, except for the immediate vicinity of zero bias for the highest
pressures, where a finer structure is revealed (see, for example, the right
panel in Fig.1).

It is seen from the Table 1 that at $P=0$ the $N_{+}/N_{e}$ ratio differs from
unity for all the samples with $N_{e}>5\cdot10^{18}$cm$^{-3}$. The
corresponding magnitudes of the characteristic electric field for tunnel
impurity ionization turned out really well above $10^{5}$ V/cm as it was
suggested. Thus, our results point out that in heavily-doped $GaAs$($Te$) the
density of the positive impurity charge inside the barrier region is
remarkably larger than the free electron density in the bulk. Fig. 2 shows an
example of the electron and charged impurity distributions in Schottky barrier
for zero and some negative biases as they have been self-consistently
calculated for values of the respective parameters corresponding to the sample
$1$ at $P=0$. This implies that not all the impurity atoms in the bulk of
heavily-doped\ $GaAs$ take part in supplying free electrons even at $P=0$. The
similar results have been obtained at $P>0$ for all the junctions, including
sample $2$.

The variation of the Schottky-barrier height $\Delta\Phi_{s}(P)$ calculated by
the above described way is presented in Fig. 3. and Fig. 4. shows the pressure dependence
of the measured and calculated zero-bias resistance of the junctions. Both
the calculated $\Delta\Phi_{s}(P)$ and the measured $\log(R_{0}(P))$ grow
slower with $P$ than it is predicted by the equality
$\Delta\Phi_{s}(P)=\Delta E_{\Gamma}(P)$ observed in \cite{PRB72} using $C$-$V$
measurements on lightly doped junctions. Moreover, there is a sharp drop of
the both quantities at the pressure $P>2.1$GPa. This drop together with the
distortion of $\sigma(V)$ curves seen in Fig. 1 can indicate some changes in
the Schottky-barrier shape due to a redistribution of impurity charge in the
depletion layer and/or a change in Fermi level pinning mechanism at the
semiconductor-metal interface. The temporal instability of the junction
characteristics observed in this range of the pressure (in particular, see
the above discussion of curves $2.2a-2.2b$ in Fig. 1) may be responsible for
different pressure dependence $\Delta\Phi_{s}(P)$ in the case of the sample $3$.

The pressure dependence of the Fermi energy of electrons in $\Gamma$-valley
for all the samples under study is shown in Fig. 5 as it was calculated from
the found $N_{e}$ values. It turned out much more regular in comparison with
$\Delta\Phi_{s}(P)$ dependencies in Fig. 3 and allows to see the strong
enhancement of the electron concentration in the case of sample $1$ at
$P=2.15$ GPa. The similar enhancement takes place in $\Delta\Phi_{s}(P)$
for this junction. It is of interest to note that just in this pressure range
the crossing of energy position of $L$- and $X$-minima should occur as can be
seen from band diagram in Fig. 5.

Preliminary conclusions may be described as follows:

1. The position of the Fermi level at metal-semiconductor interface is shifted
closer to the middle of the band gap at $N_{Te}>5\cdot10^{18}$ cm$^{-3}$
resulting in some decrease of Schottky-barrier height and its pressure
dependence.

2. After the crossing of energy position of $L$- and $X$-minima takes place the
pressure dependence of the barrier height and the tunnel junction resistance
changes its slope from positive to negative.

3. Right in the region where the band minima crossing occurs the most explicit
temporal instability of tunnel $I$-$V$ characteristics takes place.

The authors are grateful to I.M. Kotelyanskii, E.N. Mirgorodskaya, V.
Koshelets, and S.A. Kovtanyuk for valuable advices and for carrying out some
technological procedures in the preparation of the samples, to A.B. Ormont for
X-ray microanalysis, to D.K. Maude and J.C. Portal for reprints of there
articles, and to Russian Foundation for Basic Researches and INTAS 97-11475
and 96-0580 projects for the financial support.

\newpage
\begin{figure}[t]
\centerline{\includegraphics[height=10cm]{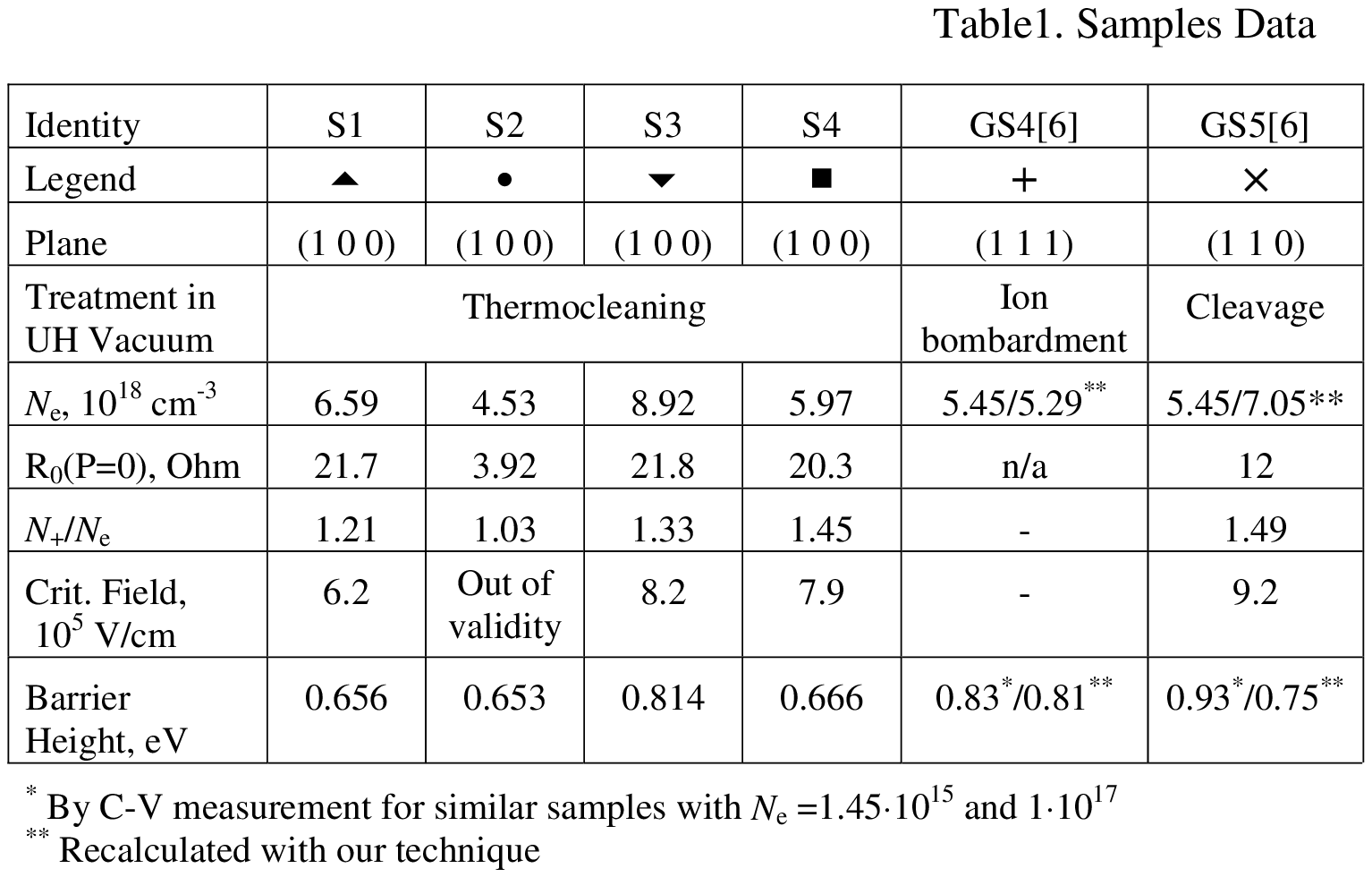}}
\label{table1}
\end{figure}
\begin{figure}[b]
\centerline{\includegraphics[height=10cm]{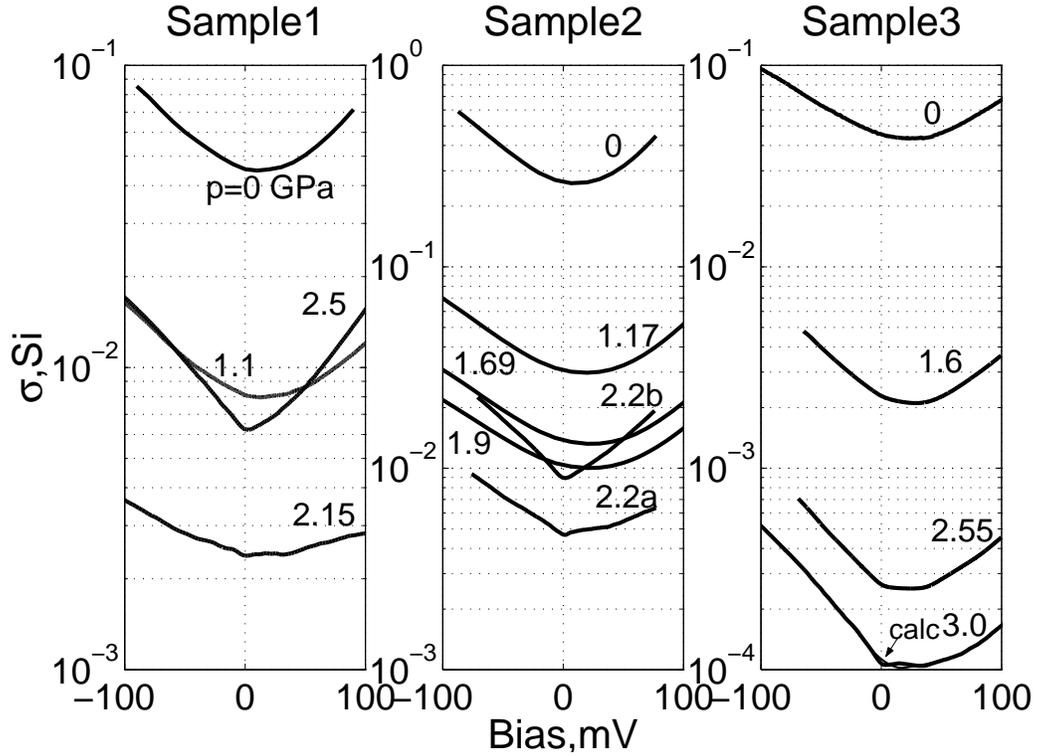}}
\caption{ Experimental $dI/dV$ curves versus bias voltage at different pressures.
Theory trends to smooth out the fine structure near zero bias occuring at the
highest pressures. The numbers over the lines represent pressure in GPa. }
\label{fig1}
\end{figure}
\newpage
\begin{figure}
\centerline{\includegraphics[height=20cm]{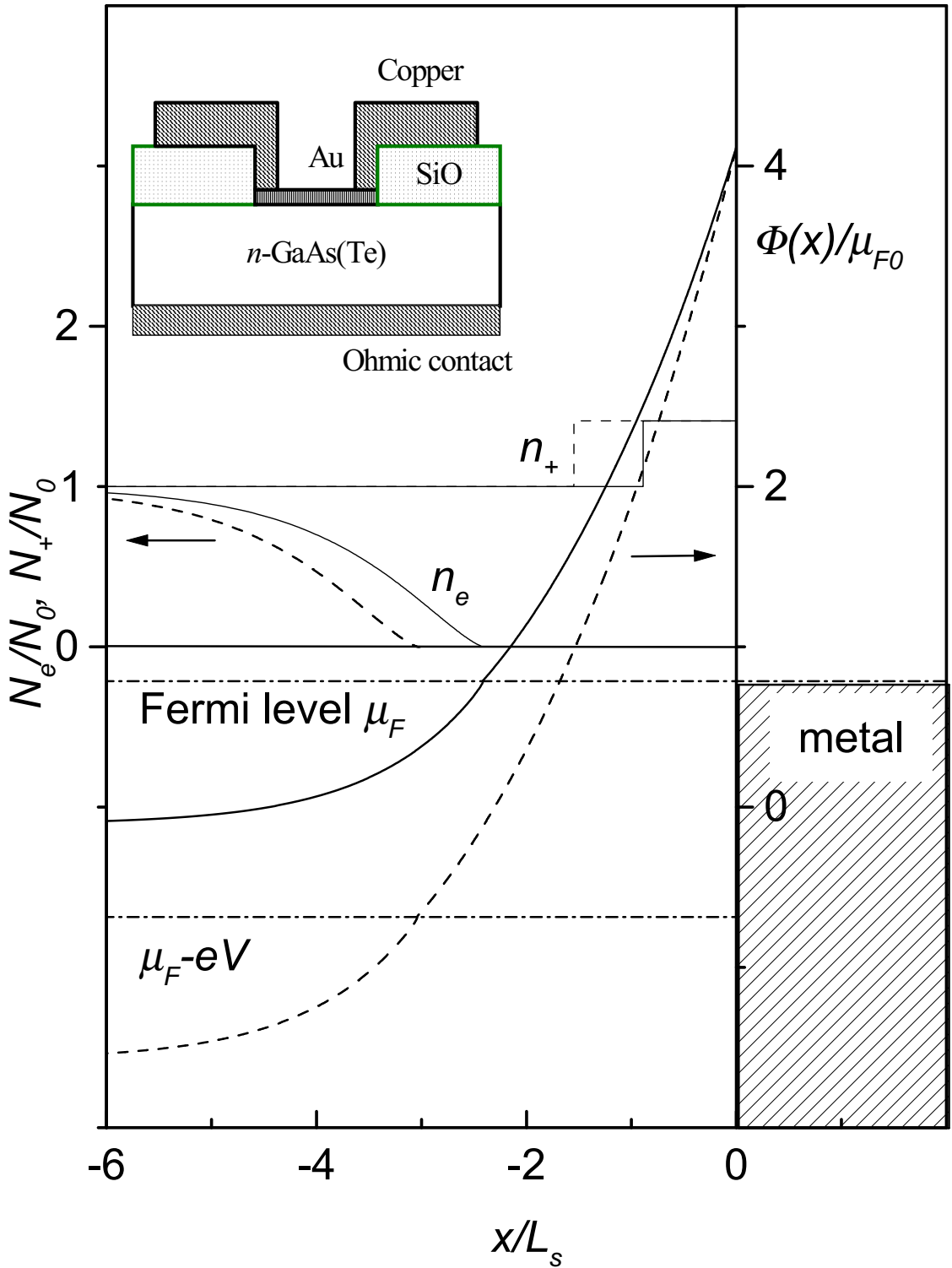}}
\caption{The reduced distributions of Coulomb potential, free $\Gamma$-valley
electrons and ionized impurity density in Schottky barrier self-consistently
calculated for two values of the bias voltage $V=0$ (solid lines) and $V=-0.3$
V (dashed lines). The density of positive charge in the depletion layer
changes abruptly at some point inside the depletion layer, reflecting the
possibly changing charge state of the impurity atoms owing to tunnel
ionization process at high enough electric field. The shape of the barrier is
obtained from the solution of the Poisson equation with the account for the
exchange interaction. The inset shows the sample assembly.}
\label{fig2}
\end{figure}

\newpage
\begin{figure}
\centerline{\includegraphics[height=10cm]{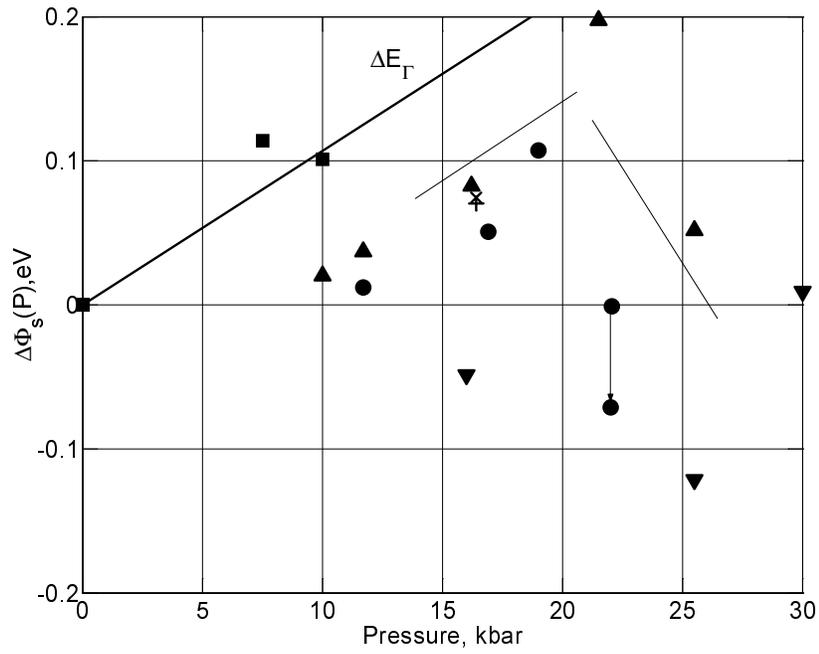}}
\caption{The variation of the barrier height $\Delta\Phi_{s}$ calculated from
the experimental $I(V)$ curves does not follow the pressure dependence of the
band gap $\Delta E_{\Gamma}$, dropping down after $2$ GPa. ({\it see Table 1 for
the legend}). }
\label{fig3}
\end{figure}

\begin{figure}
\centerline{\includegraphics[height=10cm]{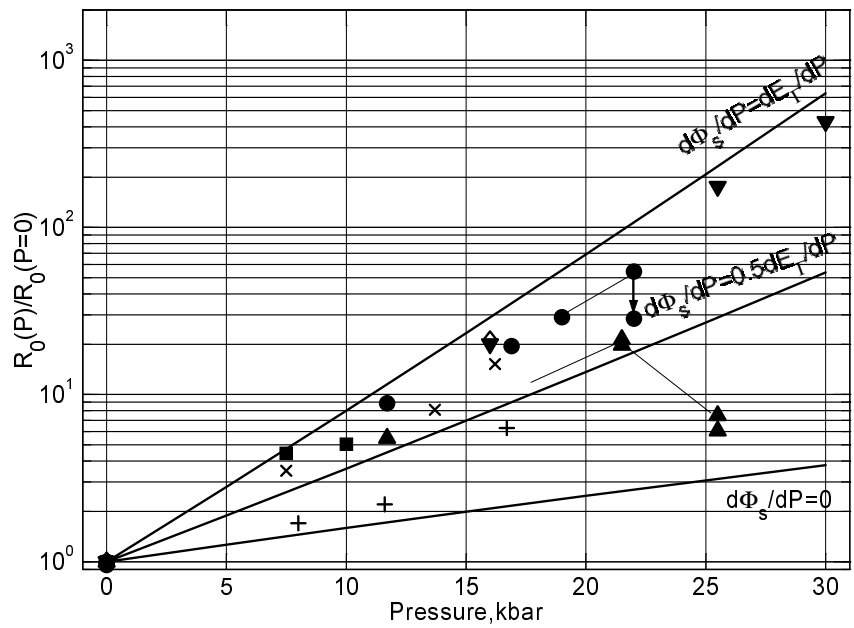}}
\caption{The measured
differential resistance at zero bias also depends on pressure differently than
it would be if one assumes $d\Phi_{s}/dP=dE_{\Gamma}/dP$ . }
\label{fig4}
\end{figure}

\newpage
\begin{figure}
\centerline{\includegraphics[height=10cm]{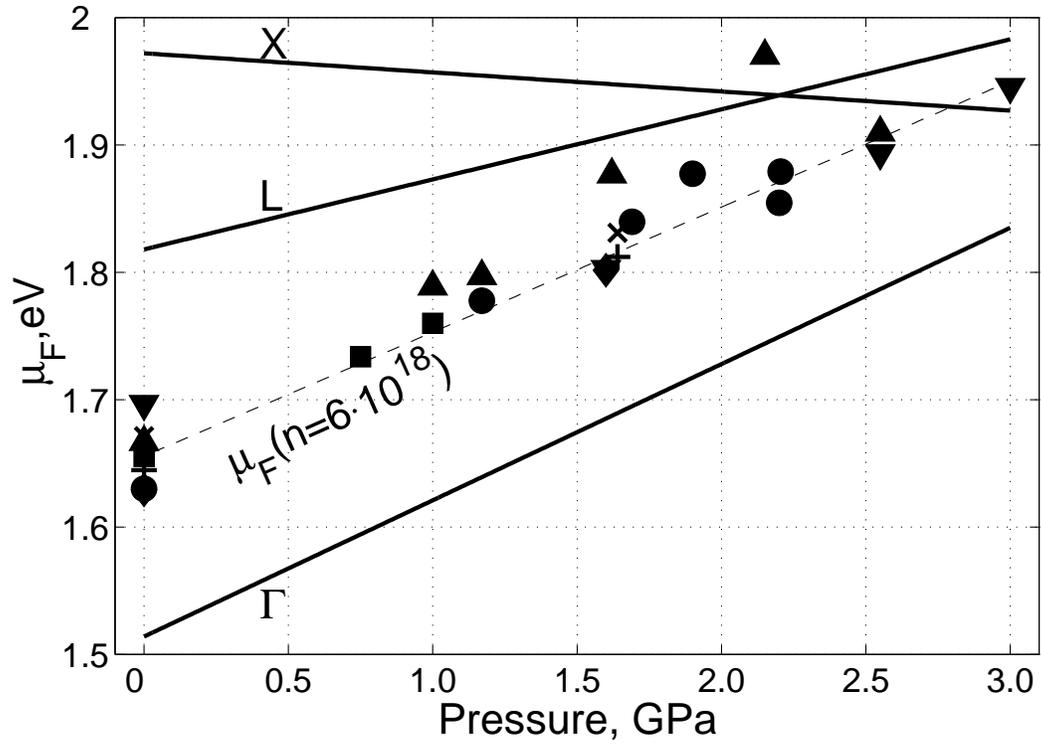}}
\caption{Fermi level calculated with the account for excange interactions for
electron densities obtained from the experiment. $L-X$ band crossing seems to
come into play. Dashed line represents the Fermi level for $N_{e}=6
\cdot10^{18}cm^{-3}$ for reference. }
\label{fig5}
\end{figure}
\end{document}